\documentclass{article}
\usepackage[accepted]{vietnam} 
\usepackage{natbib}
\usepackage{graphicx} 
\begin{document}
\twocolumn[
\title{Two class I very low-mass objects in Taurus}
\titlerunning{Two class I very low-mass objects in Taurus}
\author{C. Dang-Duc$^{1,2}$ and N. Phan-Bao$^{1}$}{cuongphysics3@gmail.com}
\address{$^{1}$Department of Physics, HCM International University-VNU, 
           Block 6, Linh Trung Ward, Thu Duc District, HCM city, 
           Viet Nam.\\
         $^{2}$Faculty of Physics and Engineering Physics, HCM University 
           of Science-VNU, 227 Nguyen Van Cu Street, District 5, HCM city,
            Viet Nam. }

\vskip 0.5cm
 
]

\begin{abstract}
We report our study of two proto-brown dwarf candidates in Taurus, 
[GKH94]~41 and IRAS~04191+1523B. Based on continuum maps at 102~GHz 
(or 2.9~mm), spectral types and the spectral energy distribution of 
both targets, we confirmed the class I evolutionary stage of 
[GKH94]~41 and IRAS~04191+1523B, and estimated the upper limit 
to the final masses to be 49$^{+56}_{-27}$~$M_{\rm J}$ and 
75$^{+40}_{-26}$~$M_{\rm J}$ for [GKH94]~41 and IRAS~04191+1523B, 
respectively. This indicates that they will likely end up as 
brown dwarfs or very low-mass stars. The existence of these class I 
very low-mass objects strongly supports the scenario that 
brown dwarfs and very low-mass stars have the same formation stages 
as low-mass stars.
\end{abstract}

\section{Introduction}
Up to date, many class II brown dwarfs (BDs) have been discovered in 
different star-forming regions. Observations of statistical properties of 
these class II BDs have strongly supported the scenario that BDs form like 
low-mass stars (see \citealt{luhman12} and references therein). However, it 
remains unclear how the BD formation process occurs at the earliest 
evolutionary stages such as BD core, class 0 and class I, which contain key 
pieces to fully understand the BD formation mechanism. So far, a few class 
0/I proto-BD candidates have been identified 
(e.g., \citealt{bourke06,lee13,palau14,morata15,liu16}) 
but only two objects of them, L328-IRS (\citealt{lee13}) and IC 348-SMM2E 
(\citealt{palau14}), have been classified as class 0 BDs with estimated final 
masses below the substellar boundary. The first pre-BD core that has been 
identified so far is Oph B-11 in rho Ophiuchi (\citealt{andre12}).

In this paper, we present our identification of two confirmed class I 
very low-mass (VLM) objects in Taurus, [GKH94]~41 and IRAS~04191+1523B.

\section{Sample and observational data}
We selected two class I proto-BD candidates, [GKH94]~41 and IRAS~04191+1523B, 
in the list of 352 members of Taurus published in \citealt{luhman10}.

[GKH94]~41 has an estimated spectral type of M7.5$\pm$1.5 (\citealt{luhman09}). 
The source was classified as a class I or very young class II object. 
\citealt{furlan11} then classified [GKH94]~41 as a class II 
object because its dereddened SED is similar to the SED of a typical T~Tauri 
star.

IRAS~04191+1523B is the secondary component of a 6.1$''$ binary 
(\citealt{duchene04}). The source has a spectral type of M6$-$M8 
(\citealt{luhman10}). In this paper, we adopted a spectral type of M7.0$\pm$1.0 
for the object. IRAS~04191+1523B was classified as a class I object in 
\citealt{luhman10}.

Observational data at infrared, submm and mm wavelengths of the targets 
available in the literature are listed in Table~\ref{t1}.

We searched for millimeter continuum observations of two targets in the 
Combined Array for Research in Millimeter-wave Astronomy (CARMA) data 
archive. [GKH94]~41 and IRAS~04191+1523B were observed on 2012 December 25 at 
102~GHz frequency (or 2.9~mm wavelength). We then reduced the continuum data 
using the MIRIAD package adapted for CARMA.

\section{Results and discussion}
\begin{figure}
\vskip -2.5cm
\centering
$\begin{array}{cc}
\hspace{-1.cm}
\includegraphics[width=7cm,angle=-90]{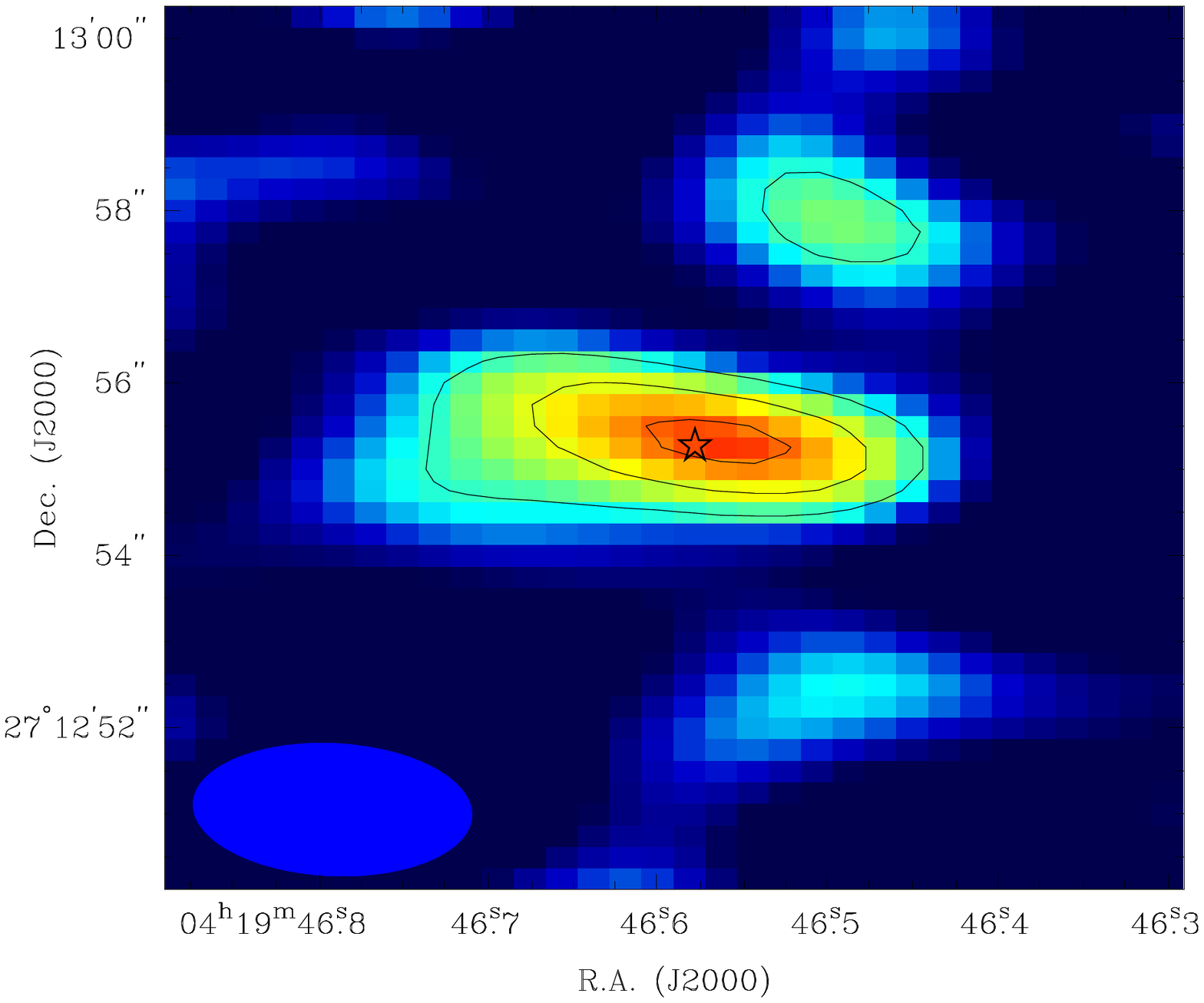} &
\hspace{-1.5cm}
\includegraphics[width=7cm,angle=-90]{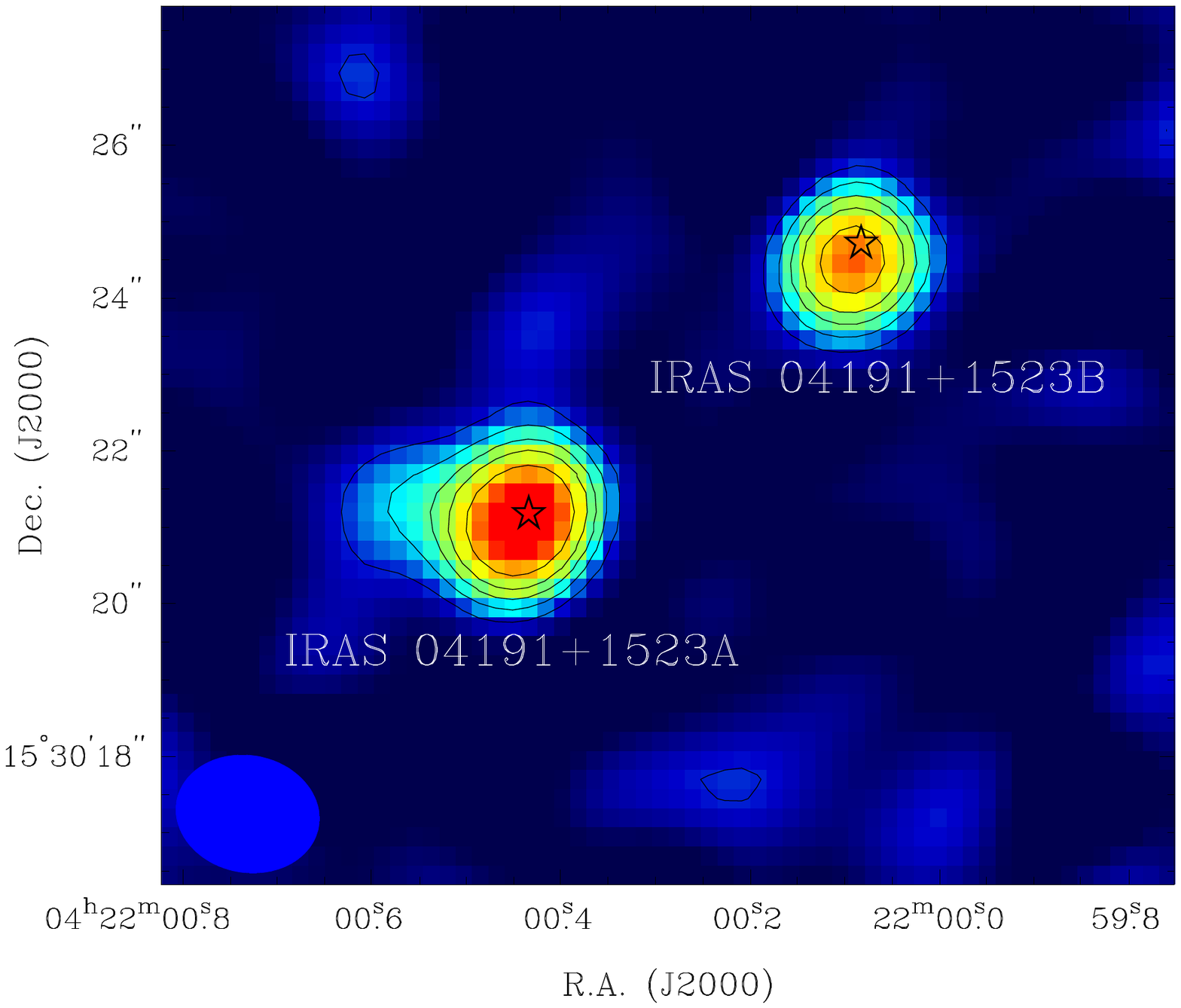} 
\end{array}$
\vskip -1.5cm
\caption{ \it {\bf Left panel:} Continuum map at 102~GHz of [GKH94]~41. 
The star symbol shows the 2MASS near-infrared position of [GKH94]~41. The 
contours are -3, 3, 5 and 7 times the rms of 0.2~mJy~beam$^{-1}$. The 
synthesized beam is shown in the bottom left corner. {\bf Right panel:} 
Continuum map at 102~GHz of IRAS~04191+1523A and IRAS~04191+1523B. The star 
symbols show the 2MASS near-infrared positions of the components. The 
contours are -3, 3, 5, 7, 9 and 12 times the rms of 0.4~mJy~beam$^{-1}$.The 
synthesized beam is shown in the bottom left corner.
}
\label{figure1}
\end{figure}

The continuum maps of [GKH94]~41 and IRAS~04191+1523B are shown in 
Figure~\ref{figure1}. Using the Gaussian fitting, we measured the integrated 
fluxes of [GKH94]~41, IRAS~04191+1523A and IRAS~04191+1523B to be 
2.5$\pm$0.2~mJy, 9.5$\pm$0.4~mJy and 5.6$\pm$0.4~mJy (see Table~\ref{t2}), 
respectively. The deconvolved sizes of the continuum emission from [GKH94]~41 
and IRAS~04191+1523A are listed in Table~\ref{t2}. For the case of 
IRAS~04191+1523B, the 2D Gaussian fitting was not able to deconvolve the 
emission, which appears to be a point source with the current spatial 
resolution.
%

%
\begin{table*}
\vskip -1.5cm
  \caption{Photometry for [GKH94]~41 and IRAS~04191+1523AB}
\label{t1}
  $$
 \begin{tabular}{ccccc}
   \hline 
   \hline
   \noalign{\smallskip}
Source            & Wavelength & Flux      & Error     & References \\
                  & ($\mu$m)   & (mJy)     & (mJy)     &            \\
\hline
[GKH94]~41        & 1.65       & 2.5       & 0.1       &(\citealt{cutri03})\\
                  & 2.17       & 11.7      & 0.3       &(\citealt{cutri03})\\
                  & 3.4        & 15        & 0.4       &(\citealt{wright10})\\
                  & 3.6        & 27.3      & 0.5       &(\citealt{luhman10})\\
                  & 4.5        & 34.2      & 0.7       &(\citealt{luhman10})\\
                  & 4.6        & 30.9      & 0.7       &(\citealt{wright10}) \\
                  & 5.8        & 41.8      & 0.9       &(\citealt{luhman10}) \\
                  & 8          & 37.5      & 0.9       &(\citealt{luhman10}) \\
                  & 12         & 46        & 1         &(\citealt{wright10}) \\
                  & 22         & 196       & 5         &(\citealt{wright10}) \\
                  & 24         & 172       & 10        &(\citealt{harvey12}) \\
                  & 70         & 269       & 5         &(\citealt{bulger14}) \\
                  & 160        & 279       & 66        &(\citealt{bulger14}) \\
                  & 2940       & 2.5       & 0.2       &(\citealt{dang-duc16})\\
   \hline
IRAS~04191+1523AB & 70         & 7002      & 13        &(\citealt{bulger14})\\
                  & 160        & 8884      & 232       &(\citealt{bulger14})\\
                  & 450        & 3940      & 220      &(\citealt{francesco08})\\
                  & 850        & 1380      & 20      &(\citealt{francesco08}) \\
                  & 1300       & 110       & 7         &(\citealt{motte01}) \\
                  & 2940       & 9.5$^{a}$ & 0.4$^{a}$ &(\citealt{dang-duc16})\\
                  &            & 5.6$^{b}$ & 0.4$^{b}$ &(\citealt{dang-duc16})\\
   \hline
   \end{tabular}
   $$
\begin{list}{}{}
  \item[]
{\bf Note:} \\
$^{a}$: the flux and its error for IRAS~04191+1523A. \\
$^{b}$: the flux and its error for IRAS~04191+1523B.
\end{list}
\end{table*}
%

%
%

\subsection{Determination of the evolutionary stages of [GKH94]~41 and 
IRAS~04191+1523B}
For the case of [GKH94]~41, the deconvolved major axis of the emission from 
the source is about 2.9$''$ (see Table~\ref{t2}) or about 400~AU in length at 
a distance of 140 pc of Taurus. This size of the emission is considerably 
larger than the typical size of disks around class II BDs (140$-$280~AU, 
e.g., \citealt{ricci14}). This suggests that the compact structure associated 
with the source is dominated by an envelope. [GKH94]~41 is therefore a class 
I object. Using the observational data in Table~\ref{t1}, we also estimated 
the bolometric temperature of [GKH94]~41 to be $\sim$460 K. This value 
indicates that the source is in the late class I evolutionary stage 
(300$-$650~K, \citealt{enoch09}). 

For IRAS~04191+1523B, it was classified as a class I object in 
\citealt{luhman10}. Our continuum map 
(see Figure~\ref{figure1}) spatially resolves the binary into two components with a separation of 6.1$''$, which is consistent with the previous value as 
reported in \citealt{duchene04}. Both components are 
associated with envelopes. This therefore confirms the Luhman et al. 
classification of the source.
%

%
\begin{table*}
  \caption{Parameters of the continuum regions around GKH94]~41, 
IRAS~04191+1523A, and IRAS~04191+1523B}
\label{t2}
  $$
 \begin{tabular}{lcccc}
   \hline 
   \hline
   \noalign{\smallskip}
Source           &  Deconvolved      & Deconvolved       &  Flux       \\
                 & major axis ($''$) & minor axis ($''$) &  (mJy)      \\
\hline
[GKH94]~41       & $2.9\pm0.1$       & $0.3\pm0.1$       & $2.5\pm0.2$ \\
IRAS~04191+1523A & $1.4\pm0.1$       & $0.3\pm0.1$       & $9.5\pm0.4$ \\
IRAS~04191+1523B & $1.8\pm0.1$$^{a}$ & $1.6\pm0.1$$^{a}$ & $5.6\pm0.4$ \\
   \hline
   \end{tabular}
   $$
\begin{list}{}{}
  \item[]
{\bf Note:} $^{a}$: Undeconvolved values.
\end{list}
\end{table*}
%

\subsection{Estimate of the final masses of [GKH94]~41 and IRAS~04191+1523B}
To determine the substellar nature of [GKH94]~41 and IRAS~04191+1523B, 
we estimate the upper limit to the final masses of these objects. Since the 
accretion mass that is added to the central object should be lower than the 
mass of the envelope associated with the object. Therefore, the upper limit 
to the final mass of a stellar object can be determined from the current mass 
of the central object and the mass of its envelope.

\subsubsection{[GKH94]~41: a class I BD}
Assuming an age of 1~Myr for [GKH94]~41 and using the effective temperature 
versus mass relation of the DUSTY model (\citealt{chabrier00}), we estimated 
the current mass of [GKH94]~41. A spectral type of M7.5 gives its effective 
temperature of about 2795~K (\citealt{luhman03}). According to the DUSTY model, 
this temperature value corresponds to a mass of 41~$M_{\rm J}$. If the 
uncertainty of 1.5 subclasses in the spectral type is taken into account, 
then the mass of the object is in the range of 14$-$97~$M_{\rm J}$. 

To estimate the envelope mass of [GKH94]~41, we searched for the best fit of 
a modified blackbody to the SED of the source with fluxes from 70~$\mu$m to 
mm wavelengths (Table~\ref{t1}) as done for IC348-SMM2E 
(see \citealt{palau14}). 
The best fit (Figure~\ref{figure2}) gives dust temperature $T_{\rm d}$ = 34~K, 
dust emissivity $\beta$ = 0.4, and envelope mass 
$M_{\rm env}$ = 2~$M_{\rm J}$. Our estimated dust emissivity is considerably 
smaller than the typical value of 1.4 for class I objects in Taurus 
(\citealt{chandler98}). The discrepancy is probably due to our best fit 
obtained with only three data points available for the source. If we use 
$\beta$ = 1.4 and $T_{\rm d}$ = 34~K to estimate the envelop mass of 
[GKH94]~41 directly from the flux at 2.9~mm, we obtain an envelope mass of 
8~$M_{\rm J}$. We then adopted an envelope mass of 8~$M_{\rm J}$ for 
[GKH94]~41.

With the current mass of 41~$M_{\rm J}$ and the envelope mass of 
8~$M_{\rm J}$, the upper limit to the final mass of [GKH94]~41 is therefore 
49~$M_{\rm J}$. If we include the spectral type uncertainty of the central 
object, the upper limit is 49$^{+56}_{-27}$~$M_{\rm J}$. This indicates 
that [GKH94]~41 will very likely end up with a substellar mass.	

\subsubsection{IRAS~04191+1523B: a class I VLM object}
The final mass of IRAS~04191+1523B was also estimated using the same steps as 
done for [GKH94]~41. A spectral type of M7.5 of IRAS~04191+1523B gives an 
effective temperature of about 2880~K. Based on the DUSTY model, this 
temperature value corresponds to a mass of 57~$M_{\rm J}$. If we include the 
uncertainty of 1.0 subclass in its spectral type, a possible mass range for 
the object will be 31$-$97~$M_{\rm J}$.

\begin{figure}
\vskip -3.cm
\centering
$\begin{array}{cc}
\hspace{-1.5cm}
\includegraphics[width=7cm,angle=0]{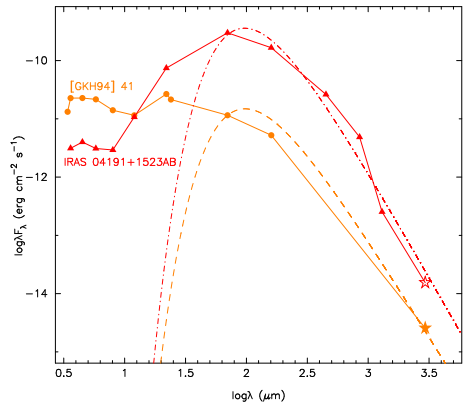} 
\hspace{-2.cm}
\end{array}$
\caption{ \it SEDs of [GKH94]~41 (brown line) and IRAS~04191+1523AB 
(red line). The solid and open star symbols show the fluxes at 2.9~mm 
(or 102~GHz) of [GKH94]~41 and~IRAS 04191+1523AB, respectively. The best fit 
for a modified blackbody of the dust envelope of [GKH94]~41 
(brown dashed-line) and IRAS~04191+1523AB (red dash-dotted line) is also 
shown.
}
\label{figure2}
\end{figure}

Because the fluxes at infrared and submm wavelengths (from 70~$\mu$m to 1.3~mm)
are unresolved for components A and B (Table~\ref{t1}). We therefore estimated 
the total mass of the envelopes associated with the two components. We then 
obtained the best fit for IRAS~04191+1523AB (Figure~\ref{figure2}) with 
$T_{\rm d}$~=~32~K, $\beta$ = 0.7, and $M_{\rm env}$ = 33~$M_{\rm J}$. Since the 
fluxes at 2.9~mm were resolved for components A and B. Therefore, the envelope 
mass of each component could be estimated directly from the 2.9~mm fluxes. If we 
use $\beta$ = 1.4 (\citealt{chandler98}) and $T_{\rm d}$ = 32~K, we obtain 
envelope masses of 31~$M_{\rm J}$ for IRAS~04191+1523A and 18~$M_{\rm J}$ 
for IRAS~04191+1523B, respectively. We then adopted an envelope mass of 
18~$M_{\rm J}$ for component B.

With the current mass of 57~$M_{\rm J}$ and the envelope mass of 
18~$M_{\rm J}$, the upper limit to the final mass of IRAS~04191+1523B is 
75~$M_{\rm J}$. If the uncertainty in its spectral type is taken into 
account, the upper limit is 75$^{+40}_{-26}$~$M_{\rm J}$. This suggests that 
IRAS~04191+1523B will end up as a BD or VLM star.
%

\section{Conclusion}
We report here our study of two class I BD candidates in Taurus, [GKH94]~41 
and IRAS~04191+1523B. Our mass estimates indicate that they will end up as 
BDs or VLM stars. The existence of these class I VLM objects together with 
two class 0 BDs and one pre-BD core as previously reported in the literature 
has demonstrated that BDs and low-mass stars have similar evolutionary tracks.

\section*{Acknowledgments}
This research is funded by Vietnam National Foundation for Science and 
Technology Development (NAFOSTED) under grant number 103.99-2015.108. 
C.D.-D. gratefully acknowledges financial support from the Local 
Organizing Committee of the Star Formation in Different Environments 
Conference held in ICISE, Quy Nhon, Vietnam on July 25-29th, 2016. 
Support for CARMA 
construction was derived from the Gordon and Betty Moore Foundation, 
the Kenneth T. and Eileen L. Norris Foundation, the James S. McDonnell 
Foundation, the Associates of the California Institute of Technology, 
the University of Chicago, the states of California, Illinois, and 
Maryland, and the National Science Foundation. Ongoing CARMA development 
and operations are supported by the National Science Foundation under a 
cooperative agreement, and by the CARMA partner universities.
%



\end{document}